\documentclass[sigconf]{acmart}
\usepackage[english]{babel}
\usepackage{blindtext}
\usepackage{mwe}
\usepackage{subcaption}
\usepackage{graphicx}

\renewcommand\footnotetextcopyrightpermission[1]{}
\setcopyright{none}
\begin{document}
\title[Unveiling Internet Censorship]{Unveiling Internet Censorship: Analysing the Impact of Nation States' Content Control Efforts on Internet Architecture and Routing Patterns}

\author{Joshua Levett}
\email{joshua.levett@york.ac.uk}
\orcid{0000-0002-6197-0241}
\affiliation{
  \institution{University of York}
  \country{United Kingdom}
}

\author{Vassilios Vassilakis}
\email{vasileios.vasilakis@york.ac.uk}
\orcid{0000-0003-4902-8226}
\affiliation{
  \institution{University of York}
  \country{United Kingdom}
}

\author{Poonam Yadav}
\email{poonam.yadav@york.ac.uk}
\orcid{0000-0003-0169-0704}
\affiliation{
  \institution{University of York}
  \country{United Kingdom}
}

\begin{abstract}
Heightened interest from nation states to perform content censorship make it evermore critical to identify the impact of censorship efforts on the Internet. We undertake a study of Internet architecture, capturing the state of Internet topology with greater completeness than existing state-of-the-art. We describe our methodology for this, including the tooling we create to collect and process data from a wide range of sources. We analyse this data to find key patterns in nation states with higher censorship, discovering a funnelling effect wherein higher Internet censorship effort is reflected in a constraining effect on a state's Internet routing architecture. However, there are a small number of nation states that do not follow this trend, for which we provide an analysis and explanation, demonstrating a relationship between geographical factors in addition to geopolitics. In summary, our work provides a deeper understanding of how these censorship measures impact the overall functioning and dynamics of the Internet.
\end{abstract}

\maketitle
\pagestyle{plain}

\section{Introduction}
\label{sec:introduction}

In this work, we present the relationship between geopolitics and the underlying network structure. The notion of a nation state's (hereafter: \emph{state} or \emph{country}) ability to regulate a sovereign Internet and undertake securitised monitoring demonstrate a strong need for the analysis of such a relationship. Previous work (particularly in the fields of geography and politics) has shown that significant changes in geopolitical relationships between nation states results in changes to route availability and the ownership of the Autonomous Systems (ASes) forming the Internet \cite{limonier_mapping_2021,douzet_digital_2023}, although this has been limited to the scope of a small number of states. We extend this work and encapsulate the wider Internet topography in our research, exploring the effect of geopolitics in defining Internet topology, both building awareness of this particular phenomenon; in particular, demonstrating the constraining impact on routing resulting from architectural structure. Furthermore, in locations where we observe constraint without strong \emph{political} intervention, we provide insight into locations in which investments in the Internet infrastructure could be important in providing valuable connectivity improvement and resilience.

Our approach collates data from the global routing table (supplemented by traceroute data), Internet resource registration data, advertised prefixes, and associated geolocations, in an effort to construct an Internet topology, and to connect the physical placement of Internet architecture with its logical equivalent. We then use a number of existing statistical approaches for graph analysis, wherein we identify a key network architectural trend among states conducting higher Internet censorship we name the \emph{funnelling effect}.

In this paper, we begin by sharing a brief outline of related work and the progress made in researching architectural trends (Section~\ref{sec:background}). We then present our approach in Section~\ref{sec:methodology}, including the function of our tooling in collecting and processing a range of sources, which we describe for each the format and processing, along with key challenges in data interpretation and combination. Following this, in Section~\ref{sec:analysis-and-results} we give an analysis of high-level architectural trends -- including in AS degree and eigencentrality -- before demonstrating the funnelling effect we find in states exerting higher effort on Internet censorship. In Section~\ref{sec:results-discussion} we then discuss a number of case studies, with examples of states following the trends we discovered alongside a small number of counter-examples, for which alternative factors result in deviation from our findings. We outline our ethical considerations in Section~\ref{sec:ethics}, followed by a summary of our work and findings in our conclusion (Section~\ref{sec:conclusion}).

\textbf{Contributions.} We make two key contributions: firstly, we present a new tool to create metadata-rich Internet topology graphs, automating the collection and processing of data from a wide variety of sources to create novel Internet state captures for fixed points in time, and at higher completeness than existing state-of-the-art, enabling further architectural study; and secondly, we present the funnelling effect discovered through our collected data, in which geopolitics (and primarily, censorship) is shown to have an architectural impact in constraining Internet routing.
\section{Background}
\label{sec:background}

Data completeness is a particularly pronounced challenge in capturing the state of the Internet topology \cite{oliveira_incompleteness_2010}. Following the demise of the IIT-CNR \emph{Isolario} project in December 2021 \cite{noauthor_isolario_2022}, the number of Internet routing data sources has decreased further, which, along with increasing adoption of private and virtual private interconnect (VPI) services, has resulted in a reduction in Internet architectural visibility.

Previous work, such as \cite{yeganeh_how_2019}, in which Yeganeh et al. utilise CAIDA's \emph{scamper} tool \cite{luckie_scamper_2010} to identify VPIs in AWS, has helped to provide additional visibility, but not at sufficient scale to identify architectural change.

In this section, we present the need for our research in the context of existing work and the significance of our contribution.

\subsection{Isolated Changes in Architecture}

Some previous work has shown the impact of significant geopolitical events on changing architecture. Most notably, this has occurred in the context of conflict or as a result of significant diplomatic shifts. In \cite{limonier_mapping_2021}, Limonier et al. highlight the impact of restrictions on route availability after the Russian territorialisation of the Crimea and Donbas regions of Ukraine. Similar analysis has been conducted for the Middle East region concerning the interconnectivity changes of the last decade \cite{douzet_digital_2023}, concluding there is minimal cooperation among Gulf countries, perceived to be resulting from the desire of nation states to exercise more control over domestic networks, however the authors argue the trends over the analysed period only partially reflect policy shifts. We hope to extend this analysis to a more technical level, but also more widely to encapsulate the complete global Internet, making delineations at the nation state level, in an effort to demonstrate architectural constraints imposed by geopolitical characteristics.

\subsection{Limited Work on Widespread Architectural Trends}

There is also little research into architectural trends at a whole-Internet level. Recent work has begun to explore this problem, including work by Bischof et al. \cite{bischof_destination_2023} on the relationship between political institutions and Internet \emph{outages} and \emph{shutdowns}, however this work is limited by considering only one potential indicator of geopolitical impact on network architecture.

Comprehensive Internet observation has shown to be a complex technological problem, largely due to challenges in observability. Our contribution makes significant improvements on previous work, with 7.1\% greater path visibility than the existing state-of-the-art \cite{wolfson_caidas_2021} through the diversification of our data sources and an improved inferencing approach.
\section{Methodology}
\label{sec:methodology}

\begin{figure}[htp]
    \centering
    \includegraphics[width=\linewidth]{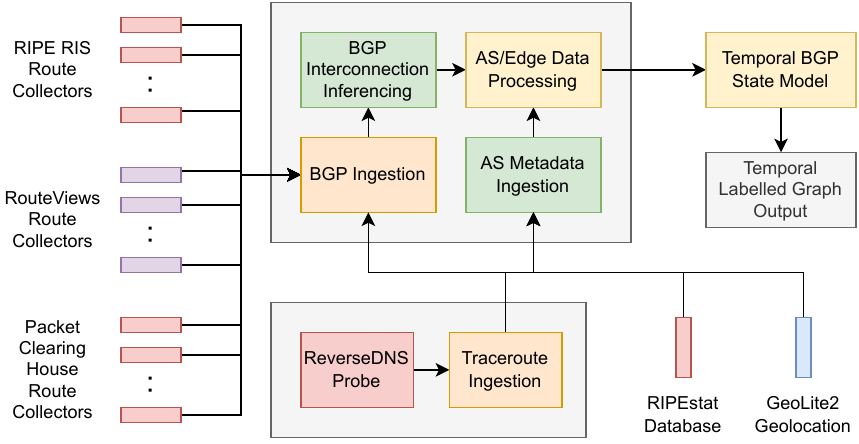}
    \caption{\textbf{System Architecture.} The figure shows the ingested data from the RIPE RIS \cite{noauthor_routing_2015}, RouteViews \cite{noauthor_university_1999} and PCH \cite{noauthor_daily_2010} route collectors, the RIPE Atlas \cite{ripe_ncc_ripe_2015} probes, RIPEstat \cite{noauthor_ripestat_2022} and GeoLite2 \cite{noauthor_geolite2_2023}. This is followed by our data processing pipeline which generates an Internet topology model underpinned by BGP inferencing, which is converted into a labelled graph object for a specified time period.}\label{diag:system-architecture}
\end{figure}

In our research, we develop a tool to collect Internet architectural data from a wide number of data sources, including a combination of database registries and Internet probes and associated processing challenges. We outline the architecture of this tool in Figure \ref{diag:system-architecture}.

The tool depends primarily on the ingested route collection feeds from each of the RIPE RIS \cite{noauthor_routing_2015}, RouteViews \cite{noauthor_university_1999} and Packet Clearing House (PCH) \cite{noauthor_daily_2010} projects. Many of these feeds are only available in compressed formats and with differing naming conventions and collection periods, which require extraction and initial processing to ensure time synchronisation. Extracted routing tables are also processed to capture both the relationship between advertised \emph{prefixes} and the claimant parent AS, which we verify where possible using historic ROA data (through data from RIPEstat \cite{noauthor_ripestat_2022}) for the specified period.

We then perform Border Gateway Protocol (BGP) connection inferencing on the data, and introduce supplementary traceroute information from RIPE Atlas \cite{ripe_ncc_ripe_2015} probes, allowing for the capture of additional peerings in the global routing table, and allowing for the validation of some of our peering inferencing.

We also ingest data from Regional Internet Registries (RIRs), largely performed using RIPEstat \cite{noauthor_ripestat_2022}, enabling the collection of historic WHOIS data and querying of the RIR databases. Finally, we ingest data from the MaxMind GeoLite2 City \cite{noauthor_geolite2_2023} geolocation database, which can assist in physically locating logical Internet resources.

To contextualise our data and demonstrate its significance, we also use data from other disciplines, which are outlined later in this section.

\subsection{Routing Data}\label{sec:routing-data}

\begin{table*}[t]
    \centering
    \begin{tabular}{lcccc} \hline 
         \textbf{Project}&  \textbf{Started} &  \textbf{Collectors} &  \textbf{\emph{Update}} & \textbf{\emph{Table Dump}} \\ \hline 
         PCH &  2010&  288\textsuperscript{$*$} &  $\geq$1 min& Daily\\
         RIPE RIS &  1999&  27\textsuperscript{$\dagger$} &  5 mins& 8 hours\\ 
         RouteViews&  1997&  45\textsuperscript{$\S$}&  15 mins& 2 hours\\
         \hline\\
    \end{tabular}
    \caption{\textbf{Route Collectors}: a comparison of routing table data availability from each project. \textsuperscript{$*$}Calculated by the number of PCH PoPs at IXPs -- not all PCH collectors publish data continually, and the available data from overall collectors changes significantly depending on the desired time range. \textsuperscript{$\dagger$}3 RIPE RIS collectors are no longer operational, and others have become live at various points since 1999. \textsuperscript{$\S$}3 RouteViews collectors are no longer operational, and others have become live at various points since 1997.}
    \label{tbl:route_collectors_and_frequency}
\end{table*}

Our key contribution in this work is to demonstrate the constraints applied to Internet routing by geopolitics, and as such, the routing data is of primary importance. Therefore, to accomplish this task, we require as complete a picture as possible of the global routing table. To achieve this, we ingest routing tables from a large number of sources, including a combination of full-feed and partial-feed sources, but from a high number of locations.

\subsubsection{Data sources}
In much published literature, data is sourced from the RouteViews and/or RIPE RIS projects. In particular, the widely shared CAIDA \cite{noauthor_caida_1970} datasets are based on a combination of the RIPE RIS and RouteViews data, supplemented with traceroute data. Previous work has also, however, highlighted inherent biases in available routing data \cite{motamedi_mapping_2019}. In the paper, the authors consider data from the two projects and highlight data bias, particularly the European/North American centrality of data collection. Indeed, in the AFRINIC and APNIC service regions, the density of route collectors falls remarkably short of that among the rest of the world \cite{motamedi_mapping_2019}. Utilised in a small number of previous works, we hope to remediate both data completeness and mitigate bias by increasing the number of BGP ingestion sources, including by including PCH data, acquired through routing table dumps at many Internet Exchange points (IXPs). We present an overview of how this data compares to more widely used data sources in Table \ref{tbl:route_collectors_and_frequency}.

\subsubsection{Data challenges}

The nature of BGP adds challenges to our route collector-centric approach:

\textbf{Attribute handling.} BGP is designed such that, without intervention, only the locally-determined optimal path to a destination is stored, rather than storing a multitude of paths \cite{limonier_mapping_2021}. This means we will collect only locally-determined optimal paths from route collector feeds. This is partially mitigated by collecting routing data from a large set of geographically distributed vantage points.

\textbf{Attribute integrity.} BGP does not provide path verification without extension, and so it is relatively trivial for BGP participants to share false paths. We attempt to mitigate this using route origin verification on the routing tables we collect using RPKI, where it has been configured.

\textbf{Vantage points.} despite being geographically distributed, the placement of vantage points is largely centred around Europe and the Americas, and therefore the completeness of our data sources, particularly in Africa and within countries of higher state censorship, is more limited \cite{sermpezis_bias_2023}.

\subsubsection{Inferring interconnections from routing tables}

Inferring the topology of the Internet through BGP has a number of well-known challenges. More than a decade ago, analysis on the same data sources highlighted the difficulty in establishing relationships due to limitations in global routing probes \cite{gregori_incompleteness_2012} -- a problem only made more difficult over time as a result of an increase in private peering and virtual private interconnect solutions \cite{yeganeh_how_2019}. Our approach, heavily utilising the \emph{as\_path} attribute, provides a foundation from which we can append paths discovered through alternate means.

Using the \emph{as\_path} attribute of each BGP record, we infer relationships between ASes based on adjacent ASes as listed in the path. For example, for any \emph{as\_path} $[x_1,..., x_{n-1}, x_n, x_{n+1}]$ where each $x$ denotes an AS, 
we consider both $x_{n-1}$ and $x_n$ as well as $x_n$ and $x_{n+1}$ to be adjacent. This enables the processing of the captured routing table and produces the set of adjacencies for a given point in time.

\subsubsection{Supplementing with traceroute data.}

In addition to our construction of the global routing table from BGP route collector feeds, we additionally collect traceroute data to both act as a validation tool for our data (as the adjacencies we collect should correlate to that of traceroute measurements), as well as to provide supplementary adjacency information, in some cases exposing \emph{private} peerings, including those unexposed in cases where our BGP feed was a partial feed.

For this purpose, we used RIPE NCC's \emph{ Atlas} tooling \cite{ripe_ncc_ripe_2015}, albeit only by monitoring traceroute results by other parties over a short 1-day period relating to the time of each BGP routing table capture. This approach is both cheaper than making new measurements, and also more widely replicable by the wider research community. Similarly to our approach with routing table data, where new routes are discovered, we supplement our existing interconnection database with the adjacencies determined through traceroute.

\subsection{Internet Registries}

Allocated by Regional Internet Registries (RIRs), either directly or delegated Local Internet Registries (LIR), Internet resources including IP address prefixes and AS numbers have a limited amount of information about their place of registration and registrant made available in RIR databases. We can use some of this information, which we believe can be fundamental to understanding the nature of allocated resources, to better understand their physical location and therefore potential geopolitical constraints.

\subsubsection{Registration country}
The registration of resources to qualifying organisations (the definition of which varies depending on a registrant's geographical location) is undertaken by five RIRs. The information collected differs in each region, but much of this information is made publicly available, albeit in differing formats. Additionally, different RIRs have different approaches to maintaining the currency of the registration data. To obtain metadata about Internet resources, we use the RIPEstat API \cite{noauthor_ripestat_2022}. Of most interest on a large scale is the registration country of each AS, which we can then use to collate topology information based on nation states.

For consistency in our work, we use the ISO 3166-1 \cite{iso_iso_2020} standardised two-letter country code (alpha-2) and the English short name.

\subsubsection{Owning organisation}
On a more limited scale, we also obtain more detailed information about the registered owner of an AS through the API, focusing primarily on its organisation. In some cases, large organisations possess multiple AS numbers. In some cases, the registering organisation information (including name) differs for those owned by the same organisation, so we also validate against CAIDA's \emph{Inferred AS to Organization Mapping Dataset} \cite{noauthor_as_2024} such that we maintain consistency across our data. As the CAIDA dataset and RIR database information are not from the same timeframe, in cases of conflicts we prefer the RIR database information.

\subsection{Resource Geolocation}

The registration location of an AS is not necessarily reflective of where it operates, and as such we produce more fine-grained information about the location of an AS based on its Points of Presence (PoPs) and the prefix it has advertised to more accurately represent the location(s) of an AS for some of our analysis.

\subsubsection{Data source}
In place of collecting our own data, we utilise the publicly available MaxMind \emph{GeoLite2 City} dataset \cite{noauthor_geolite2_2023}, which provides approximate physical location data for IP addresses, and enables us to store a local copy from which to perform IP location lookups.

The advertised accuracy of this dataset is, at the country level, claimed to be $99.8\%$, although the accuracy at a city level differs significantly between countries, with a small number of countries having correct resolutions in below $50\%$ of cases \cite{noauthor_geolocation_2023}.

\subsubsection{Utilising peer\_ip for PoP locations}
To classify the location of AS edge routers (PoPs), the locations for which we see the AS have an observable presence for observed peering, we use the \emph{peer\_ip} attribute from the collected routing tables (as described in Section \ref{sec:routing-data}). In each record, the \emph{peer\_ip} is linked to the origin AS of a BGP announcement, and therefore demonstrates activity on behalf of that AS. We perform a lookup of this IP address within the \emph{GeoLite2 City} dataset and map the geolocated country, and then add this country as an attribute of the AS in our dataset. It is important to note that it is not uncommon for one AS to have multiple countries listed in this approach (reflecting the international nature of Internet-connected networks and in particular Internet backbones, carriers, and Tier 1/2 ASes).

\subsubsection{Geolocating advertised prefixes}
We also perform some more granular analysis of the physical locations of networks, choosing to investigate the physical location of addresses within the advertised prefixes of ASes, as obtained from the collected BGP routing tables. In this case, we perform a lookup for IP addresses within an advertised prefix, and cumulatively collect the locations to propagate back to be listed as physical locations for an AS, albeit contained within a secondary attribute to that of the \emph{peer\_ip} locations. Through this approach, we can make observations about the characteristics of the ASes for which we have data, including how the observed locations for the PoPs of an AS relate to the observed locations of its IP prefix range(s).

\begin{figure}[htp]
    \centering
    \includegraphics[width=\linewidth]{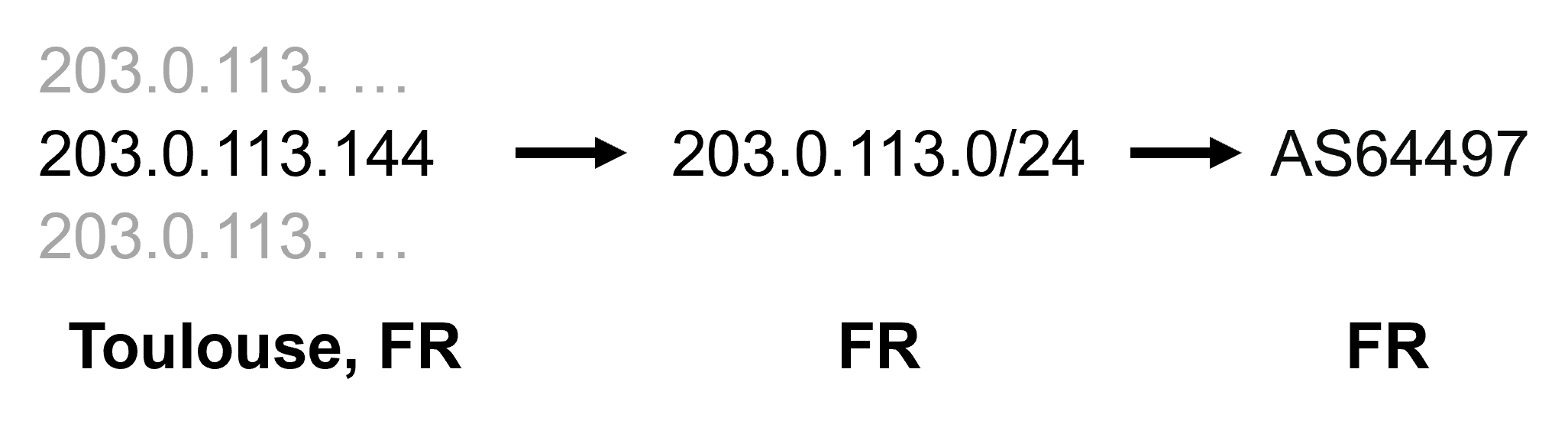}
    \caption{\textbf{Inferring the location of an AS by propagating geolocation data.} We propagate the location of IP addresses within the advertised prefixes of an AS to infer the physical location(s) of that AS. We do this both for the observed \emph{peer\_ip} in the global routing table, as well as separately for its advertised prefix ranges.}\label{fig:country-propagation}
\end{figure}

We illustrate in Figure \ref{fig:country-propagation} an overview of this approach, demonstrating the propagation of the location of one IP address within a prefix to be listed as one (of potentially many) locations for that prefix, and then similarly, as one of the locations of the advertising AS.

It is important to note that we utilise a less granular geolocation dataset to perform this analysis, connecting IP addresses only to the \emph{city} level, and then utilise only the country-level data for analysis, in an effort to minimise the risk to individuals. The ethical implications of this geolocation are discussed in Section \ref{sec:ethics}.

\subsection{Constructing a Topology}

\begin{figure}[htp]
    \centering
    \includegraphics[width=0.7\linewidth]{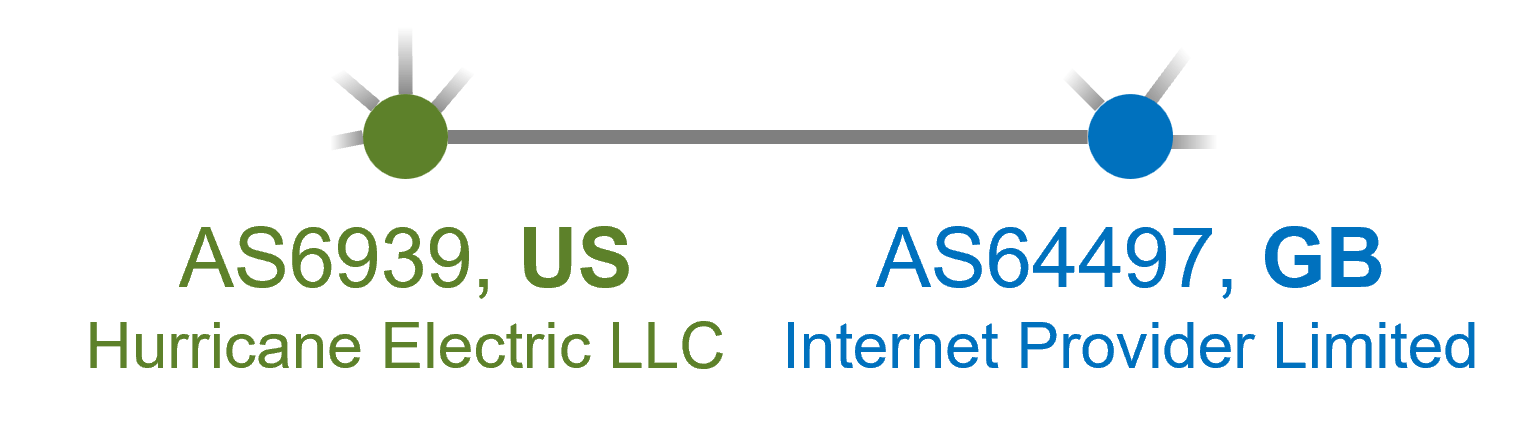}
    \caption{\textbf{Representing Internet topology in graph format.} Each AS for which we have collected data is represented as a \emph{node}, alongside attributes providing metadata (here, showing the country of registration and registered owner) and each adjacency is represented as an \emph{edge}.}\label{fig:graph-diagram}
\end{figure}

Our collected data about AS adjacency, registered location, registrar and geolocations can be used to construct a graph-based representation of the Internet's topology. We do this by collecting information from each source such that all information reflects a short temporal window, and then processing the data such that each AS becomes a \emph{node} in the graph, and each adjacency forms an \emph{edge}. In addition, we assign each country recognised by ISO 3166-1 \cite{iso_iso_2020} a unique colour. An example of this is shown in Figure \ref{fig:graph-diagram}.

Selecting data from 1st June 2023, we can in this way construct a graph of $82,593$ ASes (nodes) and $176,422$ adjacencies (edges). This is a comparative $7.1\%$ increase in observed ASes relative to CAIDA's 2023 data \cite{wolfson_caidas_2021}.

\subsection{Additional Datasets}

We apply additional sources to our work to supplement data we have collected, providing contextual data with which we perform comparisons in our analysis.

\subsubsection{Varieties of Democracy (V-Dem)}
For drawing comparisons against democratic indicators, we utilise the scoring mechanisms provided in the 2023 Varieties of Democracy (V-Dem) v13 dataset \cite{noauthor_v-dem_nodate, pemstein_v-dem_2022}, which provides a vast number of indicators for countries at annual granularity. We utilise primarily the \emph{v2mecenefi} and \emph{v2x\_polyarchy} attributes from this dataset.

\subsubsection{State-Owned ASes}
We also use the slightly outdated work from Carisimo et al. \cite{carisimo_identifying_2021} to identify state-owned ASes. The available data is no longer completely fresh, however we validate whether the registered owner in the dataset is the same as we record in our data, and if so, we assume whether the AS is majority state-owned remains accurate. This may not always be true, however we believe it is still pertinent to use this data at an overview scale for the benefit of making generalised conclusions, even if the data at a granular level may not be thoroughly correct.

\section{Analysis and Results}
\label{sec:analysis-and-results}

In previous contributions in this area, the focus has largely been on routing changes, either by highlighting connection state changes or correlating Internet shutdowns and outages with various autocracies \cite{limonier_mapping_2021, bischof_destination_2023}. Our work presents a new perspective: we highlight instead the structure of networks in respective nation states, and the commonality of network structure in different states.

\subsection{Topology Observability}

\begin{figure}[htp]
    \centering
    \includegraphics[width=\linewidth]{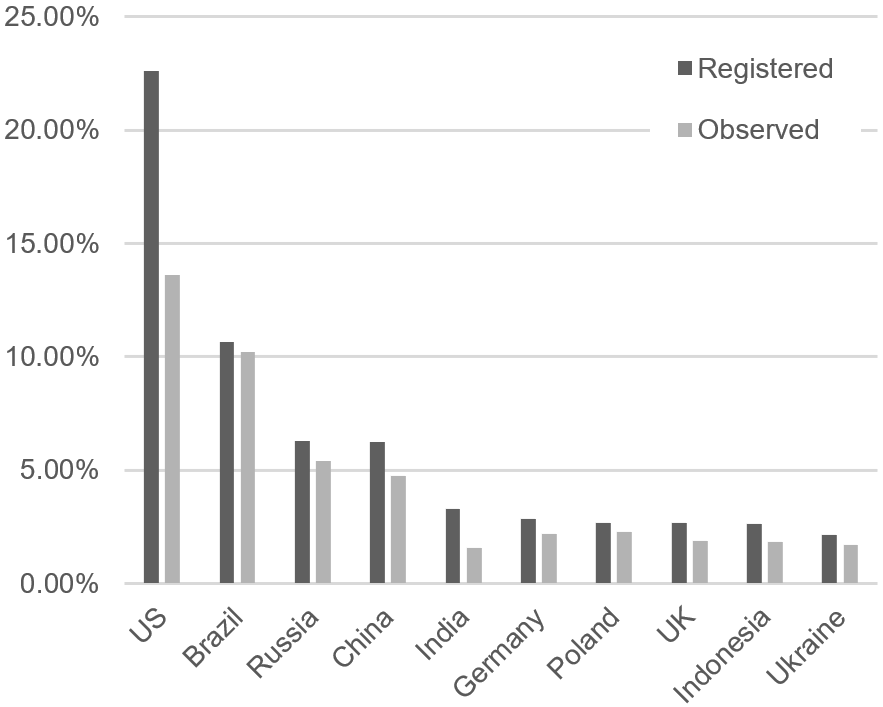}
    \caption{We observe at least one relationship for the majority of registered ASes (showing 10 highest countries by AS registration).}\label{diag:observability}
\end{figure}

We capture the state of the BGP routing table at 00:00 (UTC) on 1 June 2023, at which point we can infer at least one interconnection relationship for $71.51\%$ of registered ASes, and with an overall inferred average AS degree of $4.05$. As shown in Figure \ref{diag:observability}, we observe relationships for the vast majority of ASes in countries with highest AS registration. Notably, however, this varies by RIR, where some regions have not reclaimed unused ASNs and hence many remain parked. This may partially explain the lower level of observation in the ARIN region, and particularly the US. We also observe $544$ bogon ASes for which there are no records of registration.

\subsection{Connectivity by State}

\begin{figure}[htp]
    \centering
    \includegraphics[width=\linewidth]{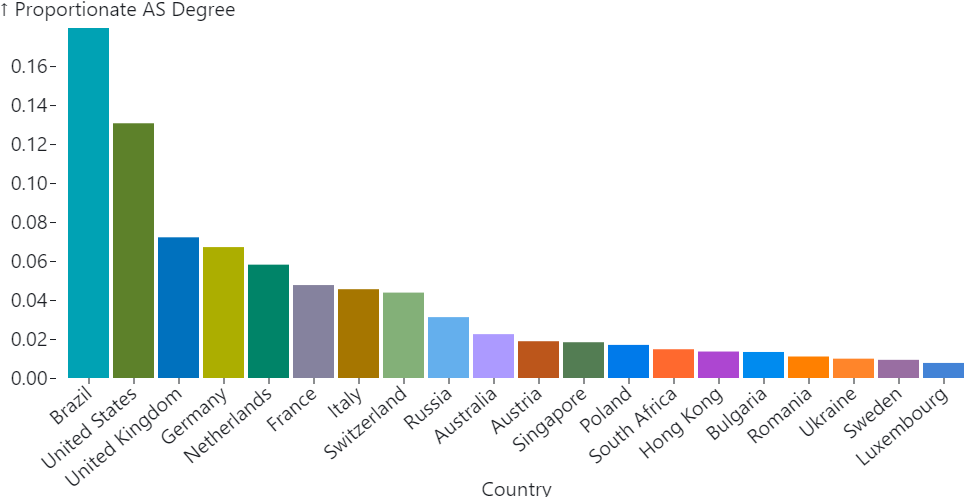}
    \caption{\textbf{Countries with the highest proportion of AS degrees (top 20).} This shows the sum of all AS degrees registered within an ISO-recognised country, divided by the sum of all AS degrees.}\label{fig:country-proportion-degree}
\end{figure}

Intuitively, ASes with a higher degree have a higher level of connectivity to other ASes. In Figure \ref{fig:country-proportion-degree}, we show that a small number of countries have the highest level of inter-AS connectivity by showing the proportion of \emph{all} inter-AS connectivity, presented by country. This approach adjusts for some countries having a particularly high number of low-degree ASes by considering the total degree for a country in place of using a state-based average.

\begin{figure}[htp]
    \centering
    \includegraphics[width=\linewidth]{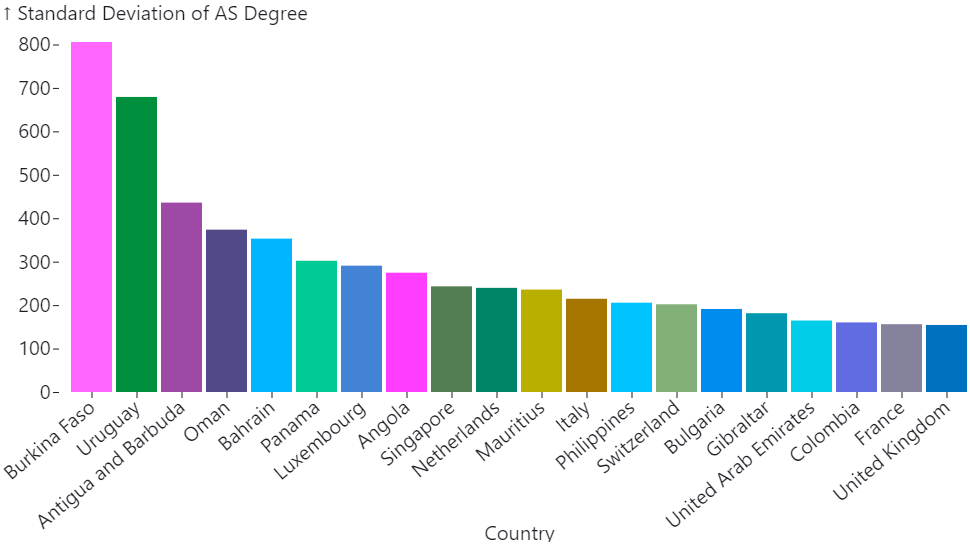}
    \caption{\textbf{Countries with the highest standard deviation in AS degree (top 20).}}\label{fig:country-std-dev}
\end{figure}

Considering AS degree also demonstrates to some extent a relationship between the degree of some ASes and the country of registration, especially when contextualised by the smaller standard deviation in many of the same countries, as shown in Figure \ref{fig:country-std-dev}. Notably, the countries showing greatest deviation have very little Internet permutation, or one primary state operator. For instance, in Uruguay, almost all connectivity is delivered by state-owned \emph{Administración Nacional de Telecomunicaciones} (ANTEL, AS6057). In Uruguay, there exists minimal state censorship and strong democratic standing \cite{noauthor_v-dem_nodate}), and therefore AS6057 has a high degree of connectivity, unlike the considerably smaller downstream ASes. The much smaller level of deviation in countries such as Ireland and the United States show that many ASes have a higher degree of connectivity.

\subsection{Eigencentrality by Country}

An alternative metric we can use to understand the \emph{influence} of an AS is that of its eigencentrality -- a measure of how `central' to the Internet a given AS is. Considering the adjacency matrix between all ASes as $\mathbb{A}$, and $\lambda$ as a constant eigenvalue we seek to maximise, we can obtain an eigenvector $x$, where each $x_n$ (eigencentrality) in $x$ is indicative of the influence of a given AS $n$ using $\lambda x = x \mathbb{A}$. We normalise $x$ such that all values $x_n$ in $x$ are less than $1$.

\begin{figure}[htp]
    \centering
    \includegraphics[width=\linewidth]{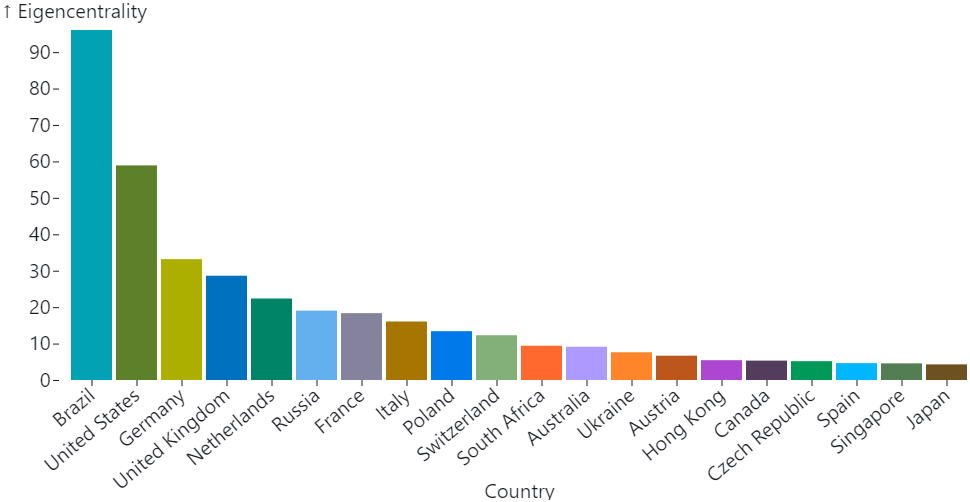}
    \caption{\textbf{Countries with the highest eigencentrality (top 20).} Plotting countries with the highest (summated) eigencentrality value, demonstrating one measure for a nation state's Internet architectural influence.}\label{fig:country-sum-eigen}
\end{figure}

Where $x_n$ for a given AS $n$ tends towards $1$, we can determine that it must therefore have more influence. In a BGP environment that was not self-correcting this would mean that the loss of $x_n \to 1$ would result in a higher loss of connectivity. We show the top 30 countries with the highest cumulative eigencentrality in Figure \ref{fig:country-sum-eigen}.

\begin{figure}[htp]
    \centering
    \includegraphics[width=\linewidth]{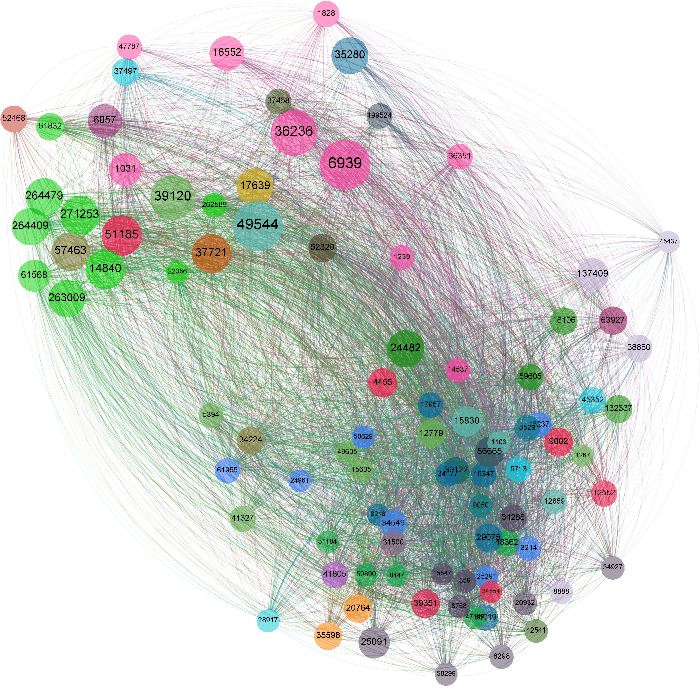}
    \caption{The ASes with highest eigencentrality (and therefore most influential) are deeply interconnected. In this figure, we scale each AS node by its eigencentrality, and propel those with lower connectivity whilst forming communities of those most interconnected.}\label{fig:influential-interconnect}
\end{figure}

When considering eigencentrality of individual ASes, we would expect a high degree of interconnectivity between ASes, which we see in Figure \ref{fig:influential-interconnect}. As follows from Figure \ref{fig:country-sum-eigen}, the highest summated eigencentrality values by country are replicated in terms of high-eigencentrality ASes.

Utilising eigencentrality as a measure reinforces similar findings to that of purely considering AS degree, which is not unexpected given eigencentrality is a metric dependent on AS adjacencies. However, it is notable that the metrics are not the same, and the eigencentrality in particular highlights the unusually high connectivity of networks in Brazil, where it is evident that many high-degree networks in Brazil must be deeply interconnected, unlike the United States where there exists a higher degree of provider tiering, reducing the direct interconnectivity between ASes and instead routing through higher-tier intermediaries. This is partially demonstrated in Figure \ref{fig:influential-interconnect} where ASes from the United States are more varied in scale between the highest eigencentrality -- Hurricane Electric LLC (AS6939) and NetActuate, Inc (AS36236), versus all others -- and the more balanced sizing of Brazilian ASes.

\subsection{Considering Foreign Neighbours}

\begin{figure}[htp]
    \centering
    \includegraphics[width=\linewidth]{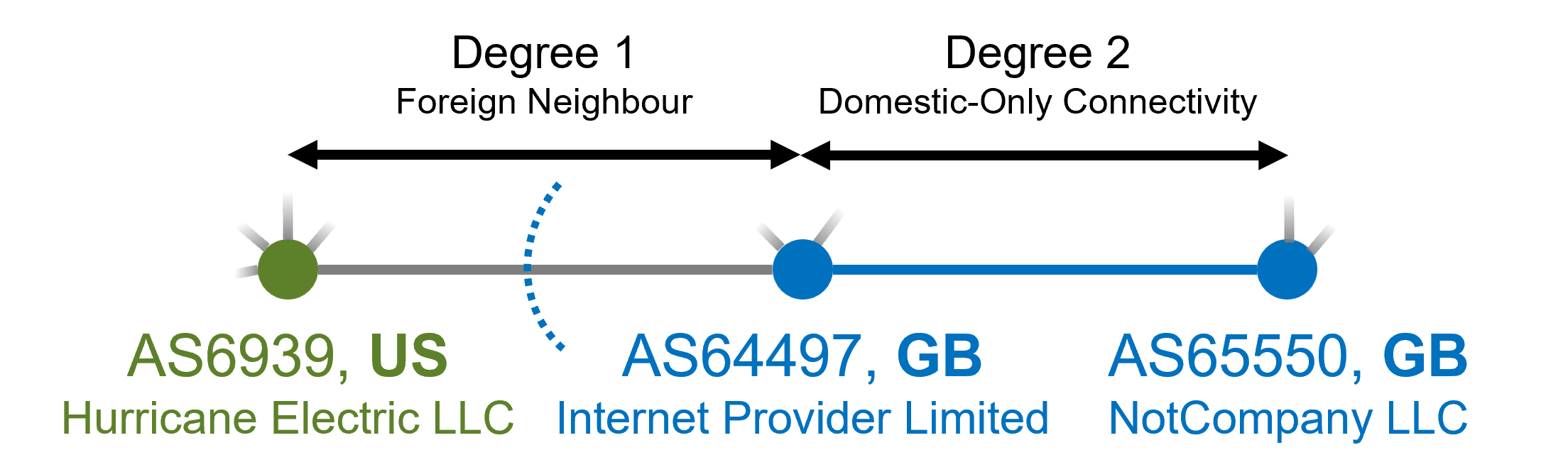}
    \caption{Defining a foreign neighbour and degrees of connection from foreign neighbours. In this diagram, we show \emph{first degree} foreign neighbour connectivity to be a direct adjacency to a foreign neighbour. Second degree connectivity is where the shortest path (regarding only hop-distance rather than path statistics) to a foreign neighbour is through only one AS.}\label{fig:foreign-neighbour}
\end{figure}

We can extend our analysis to consider the degree of non-domestic neighbours of an AS, which although providing some information about the level of a state's interconnectivity with the wider world, also leads to more detailed conclusions about access to the Internet later in this section. To deem a neighbour \emph{foreign}, we consider only whether its country of registration differs from an AS under consideration. As demonstrated in Figure \ref{fig:foreign-neighbour}, where AS64497 is under consideration, we regard AS6939 as a first degree foreign neighbour as the ASes are registered in differing countries.

Taking this more simplistic approach, although less indicative for states in which a large number of tier-1/carrier providers operate, would appear to be more indicative of international connectivity in less-connected states, as shown in the case studies presented in Section \ref{sec:results-discussion}.

\subsection{Route Funnelling}

Furthermore, we present a measurement for what we will regard as the \emph{funnelling} effect, wherein connectivity from a country is only possible by traversing a small set of ASes -- which we demonstrate are, in cases where this effect is most clearly observed, usually state-controlled.

\subsubsection{Relative downstream change}
\begin{figure*}[tp]
    \centering
    \includegraphics[width=\linewidth]{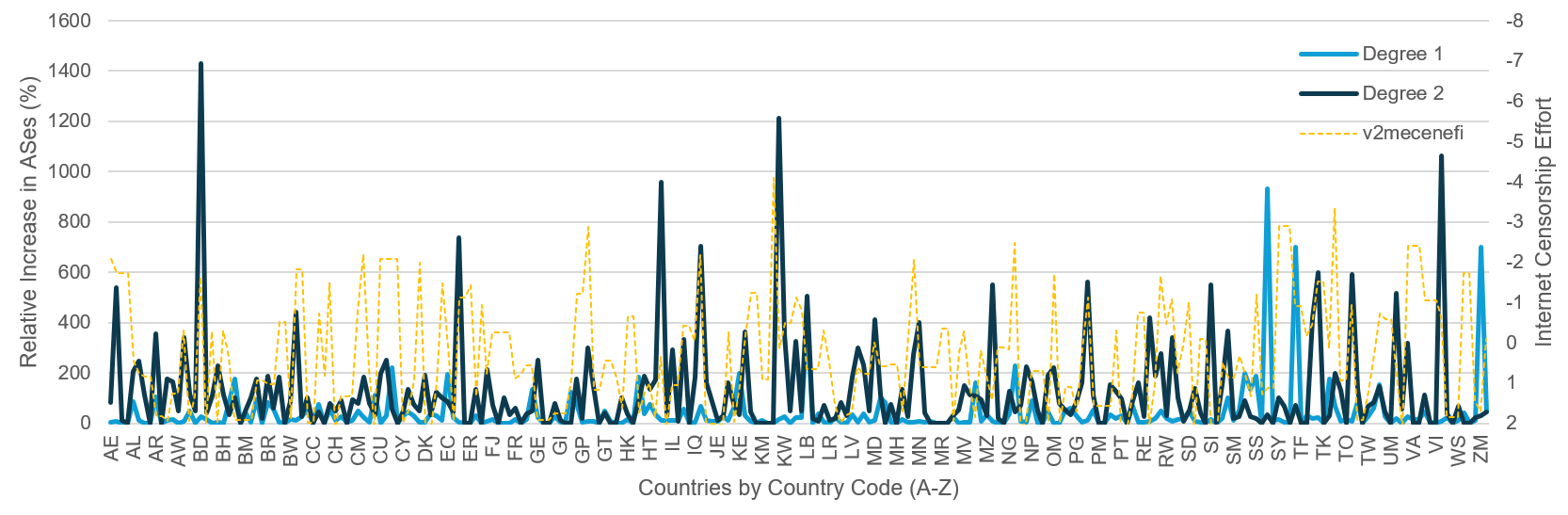}
    \caption{\textbf{Demonstrating the relationship between change the relative change in ASes at increasing distance from a foreign AS and Internet censorship effort.} In this figure, we can see that changes in Internet censorship effort (from V-Dem's \emph{v2mecenefi} data \cite{noauthor_v-dem_nodate}) correlate to significant change in the number of ASes at degree 2.}\label{fig:funnel-purechange}
\end{figure*}

We first consider the change in the number of ASes at each hop from domestic ASes with foreign neighbours. We assign the set of ASes connected but foreign to a given state as set $d_0$, and then generate a set $d_1 := \mathrm{neighbours}(d_0) \setminus d_0$ of all domestic neighbours of $d_0$ not contained in the set $d_0$. We continue this approach for each set $d_{n+1} := \mathrm{neighbours}(d_n) \setminus (d_n \bigcup d_{n-1} \bigcup ...)$ until all domestic ASes are included across all sets. Using the size of each set $d_n$, we calculate the relative increase at each hop, the first two of which are displayed in Figure \ref{fig:funnel-purechange}.

In this approach, we identify a weak correlation ($\approx -0.4$) for each of the first two hops (which we consider to be \emph{first} and \emph{second} degree to foreign neighbours).

\subsubsection{Considering the cumulative downstream burden}
In our second approach we consider the relative number of downstream ASes for each AS at the each hop in this process relative to all ASes at downstream. This particular metric eliminates the misleading skew placed on the previous metric where ASes at the first hop may have no domestic neighbours.

In this approach, we calculate for each AS with foreign neighbours the total number of domestic downstream ASes it has (where this is nonzero) and divide this by the number of domestic ASes with foreign neighbours. We then find the mean of these AS values, and repeat the approach for the next hop until the mean value for the hop reaches $0$.

\begin{figure}[htp]
    \centering
    \includegraphics[width=\linewidth]{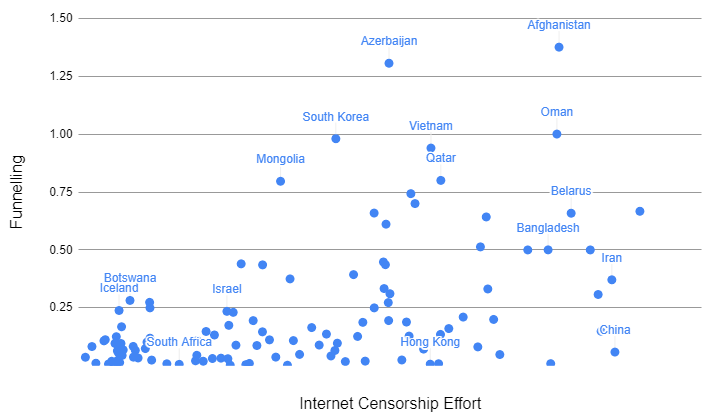}
    \caption{\textbf{Demonstrating increased levels of funnelling in countries with higher Internet censorship.} This figure shows a tight cluster of countries with low censorship and minimal funnelling, followed by increasing detection of funnelling in countries of higher censorship according to the V-Dem \emph{v2mecenefi} data \cite{noauthor_v-dem_nodate}.}\label{fig:funnel-with-censorship}
\end{figure}

For each hop, a higher value is indicative of some aspect of funnelling, wherein a small number of ASes provide the majority of downstream domestic connectivity. We therefore expect, following the pattern indicated in the previous approach, that countries employing higher Internet censorship effort will have higher mean values, particularly at the first hop. Figure \ref{fig:funnel-with-censorship} shows this to largely be true, wherein at higher levels of Internet censorship effort, the level of first hop funnelling increases substantially for a number of states.

There are a few countries with lower levels of censorship for which the level of funnelling is higher than might be anticipated. Notably for Iceland, connectivity is restricted largely by geographical placement, whereas Botswana is currently experiencing extremely rapid growth in domestic Internet infrastructure and connectivity largely resulting from state intervention and as such its relatively higher level of funnelling results from other factors than censorship. Other countries with lower censorship effort include Mongolia and South Korea, each facing geopolitical constraints to connectivity as a result of neighbouring physical countries.

\subsubsection{Comparison with eccentricity}

\begin{figure}[tp]
    \centering
    \includegraphics[width=\linewidth]{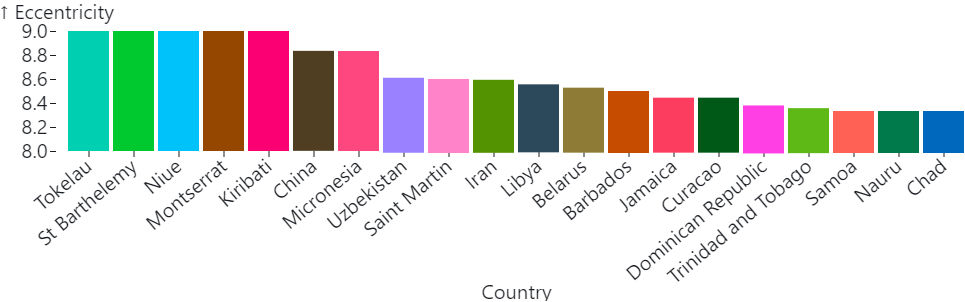}
    \caption{\textbf{Countries with the highest mean eccentricity (top 30).} In this figure, we calculate the mean eccentricity (maximal distance between ASes) for each country.}\label{fig:country-eccentricity}
\end{figure}

Eccentricity is an established metric for calculating the greatest distance between two points on a graph. We show in Figure \ref{fig:country-eccentricity} the difference between our results for the funnelling effect in each state and eccentricity, which largely depicts the level of AS downstreaming in some states.

\subsection{Graphical Representation}

There is some previous work in graphically plotting interconnectivity within the Internet. CAIDA provides a series of visualisations for Internet data, the most relevant to this work being that of the IPv4 and IPv6 AS Core by Wolfson et al. \cite{wolfson_caidas_2021}. In this work, the authors utilise BGP data captured by the RouteViews and RIPE RIS projects complemented by traceroute data sourced using the CAIDA Ark tool \cite{noauthor_archipelago_2007}. Similarly, another notable approach to Internet mapping is The Opte Project \cite{lyon_opte_2003}, which in more recent years, has used purely BGP routing tables collected from the RouteViews project \cite{noauthor_university_1999}.

In comparison to both works, however, our approach (supported by an increased pool of data sources) achieves much greater topology completeness, and is more readily replicable for data collection for an increased number of historic timepoints. Indeed, in our work we show that for the same time period, we can capture an additional $4.6\%$ ASes and more significantly, $7.1\%$ more inter-AS peerings. Additionally, our work is novel in that the focus differs; where Wolfson et al. represent the AS core, and the significance of a small number of ASes constituting a central AS `cone' and The Opte Project shows connections with effective equal weighting, we instead work to expose the increase in Internet fragmentation over time using force-directed graphical representations.

Our topology graph visualisation approach, utilising the ForceAtlas2 algorithm \cite{jacomy_forceatlas2_2014}, tightly groups highly connected nodes and repels nodes with the fewest connections. The algorithm itself is iterative, where we reach a point of convergence but never a state of completion, and thus we end iterations after reaching a state of visual convergence.

In this alternate area, there is some existing work focusing on the changes in the Internet topology graph around isolated geopolitical events. In the context of Eastern Ukraine, researchers at GEODE utilised BGP topology data to demonstrate the topological impact of geopolitical tensions and later conflict \cite{limonier_mapping_2021}. Despite also enabling geopolitical analysis, our work marks a significant improvement: our topology capturing methodology outperforms state-of-the-art by reaching increased completeness; and additionally we present our research on a vastly greater (macroscopic) scale, observing change on the overall Internet topology than localised geopolitical events.
\section{Discussion}
\label{sec:results-discussion}

Following our analysis, here we present our key findings and notable results. Fundamentally, our research has shown that a relationship exists between states exerting high Internet censorship effort and a resulting change in Internet architecture. Our novel approach to measuring this goes beyond traditional graph metrics and demonstrates a constraint on Internet routing resulting from geopolitics.

\subsection{Case Studies}

In this section we demonstrate a few examples of countries following key architectural trends based on our analysis. We provide a numerical overview of these countries in Table \ref{tbl:case-studies}.

\begin{table*}[t]
    \centering
    \begin{tabular}{lcccc} \hline 
         \textbf{State} &  \textbf{Foreign Neighbours} &  \textbf{First Degree to Foreign} &  Unique Registrants (\textdegree 1) & \textbf{Internet Censorship} \\ \hline 
         China & 1,460 & 300 & 70\textsuperscript{$\dagger$} & -2.202 \\ 
         India & 1,010 & 551 & 116 & -0.41 \\ 
         Iran & 79 & 43 & 1 & -2.176 \\
         Russia & 2,733 & 1,278 & 19 & -1.662 \\
         United Kingdom & 9,478 & 1,502 & 1,307 & 0.552 \\
         \hline\\
    \end{tabular}
    \caption{\textbf{International connectivity comparison.} We display summary statistics for each country's number of foreign AS neighbours, the number of ASes with first degree connectivity to foreign neighbours, the number of registrants delivering such connectivity, and then the V-Dem dataset's summary statistic for \emph{Internet censorship effort} (v2mecenfi) \cite{noauthor_v-dem_nodate} wherein lower scores are indicative of diminishing access to the Internet, with mean access (base) given at $0$. \textsuperscript{$\dagger$}This is not entirely representative, as almost all such registrant organisations are majority-owned by the Chinese state \cite{carisimo_identifying_2021}.}
    \label{tbl:case-studies}\vspace{-5mm}
\end{table*}

\subsubsection{United Kingdom}

\begin{figure}[htp]
    \centering
    \includegraphics[width=\linewidth]{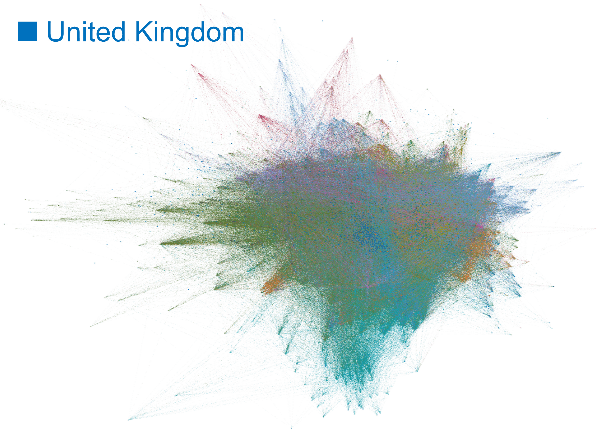}
    \caption{\textbf{International connectivity of the United Kingdom.} The figure shows the United Kingdom and its first degree of foreign neighbours. In this case, the domestic network is almost entirely encapsulated within the dense graph of its neighbours, demonstrating strong interconnectivity.}\label{country:united-kingdom}
\end{figure}

In this first example, we present a state with a high degree of international connectivity and diversity of connectivity provider. When considering all ASes registered within the United Kingdom, we calculate it has $9,478$ non-domestic neighbours reached from $1,502$ foreign-facing domestic ASes. This notably high degree is demonstrated visually in the force-directed representation shown in Figure \ref{country:united-kingdom}, wherein the state is represented almost indistinguishably in the centre of the graph and amongst its foreign neighbours.

Internet infrastructure provider consolidation is a distinct discussion not covered in this work, however we identify $1,307$ registered AS owners with first degree foreign connectivity -- $71.0\%$ of all ASes registered in the state. With high levels of polyarchy and low Internet and media censorship \cite{noauthor_v-dem_nodate}, our funnelling analysis shows little relative downstream AS change where each downstream hop has fewer ASes, and we observe particularly low levels with the cumulative downstream burden approach -- $93\%$ below the lowest quartile value.

\subsubsection{Iran}

\begin{figure}[htp]
    \centering
    \includegraphics[width=0.8\linewidth]{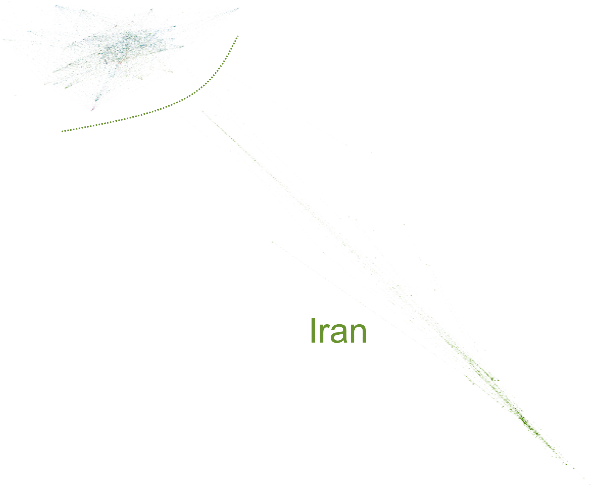}
    \caption{\textbf{International connectivity of Iran.} The figure shows Iran and its first degree of foreign neighbours. Iran has a clear disconnect between its foreign neighbours, with only $7$ ASes providing foreign connectivity.}\label{country:iran}
    \vspace{-8mm}
\end{figure}

When we consider Iran's Internet connectivity -- a state wherein Internet censorship efforts are considerably more pronounced -- the force-directed graph in Figure \ref{country:iran} demonstrates clearly the disjunct between Iran and the wider Internet. In our data, we observe only $79$ foreign neighbours when considering ASes within the state, to which connections are only made from $43$ Iran-registered ASes.

It is also clear from the topology that, despite the $43$ ASes with foreign neighbours, almost all of Iran's connectivity is delivered through the \emph{Telecommunication Infrastructure Company} (AS49666). Thus, without considering the funnelling effect, much of the architectural restriction experienced by Iran would not be readily observable.

\subsubsection{India}

\begin{figure}[htp]
    \centering
    \includegraphics[width=\linewidth]{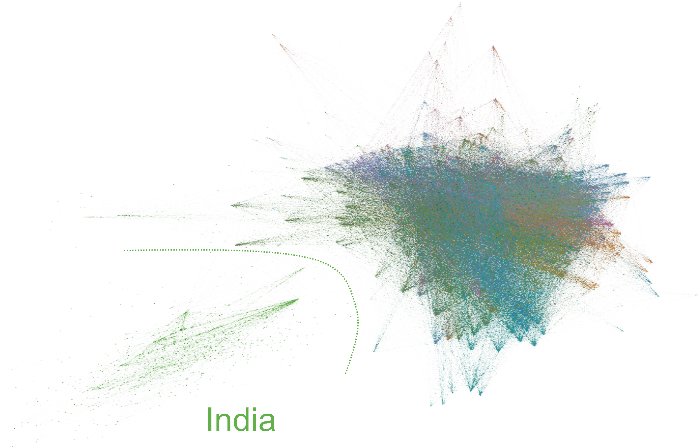}
    \caption{\textbf{International connectivity of India.} The figure shows India and its first degree of foreign neighbours, in this case showing a strong divide between its neighbours and its domestic Internet, with clear hub ASes separating the domestic Internet.}\label{country:india}
\end{figure}

In this third example, we observe the connectivity of India. As we can observe statistically as a trend in states with a notable degree of Internet censorship effort, Figure \ref{country:india} demonstrates a clear and visible separation between the domestic connectivity of many of India's ASes (of which we identify $2,651$ on 1 June 2023), and limited connectivity to foreign ASes, delivered by only $20.8\%$ of its ASes.

\begin{figure}[htp]
    \centering
    \includegraphics[width=\linewidth]{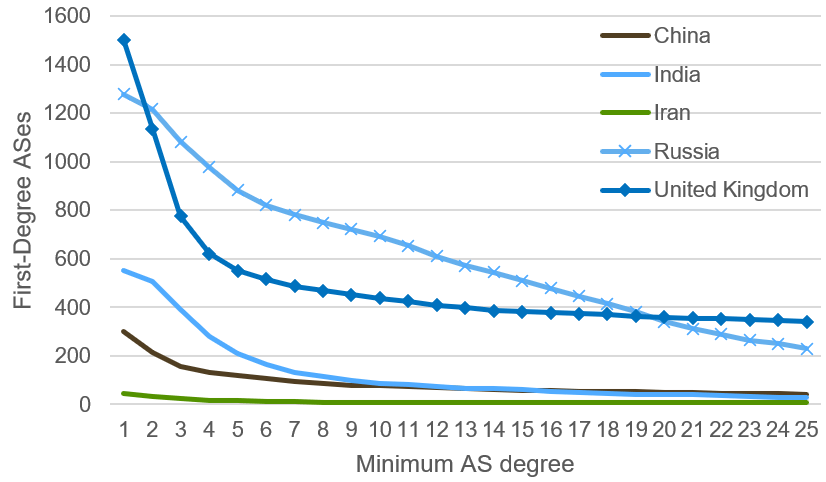}
    \caption{\textbf{Diminishing foreign connectivity when introducing a minimum AS degree threshold.} This figure shows the number of domestic ASes in each state with foreign connectivity over an increasing minimum degree threshold. We use this to show nodes with more architectural significance.}\label{fig:degree-over-threshold}
\end{figure}

If we analyse this international connectivity further, however, we can identify that where we consider only ASes with a degree above a threshold $x$, the number of ASes delivering first degree foreign connectivity reduces to $\approx80$ where $x\geq9$, as shown in Figure \ref{fig:degree-over-threshold}.

The impact of such architectural phenomena is measurable when using our approach to identifying funnelling. Despite a relatively low cumulative downstream burden metric of $0.025$ -- equivalent to the lower quartile, but likely due purely to the relatively equal distribution between a wider number of the funnel nodes -- India has a significant downstream change with a $43\%$ decrease in ASes from foreign ASes to the first domestic hop, followed by a $334\%$ increase. This is supported visually in the force-directed graph shown in Figure \ref{country:india}.

\subsubsection{Russia}

Finally, we present a counterexample to our work. Russia has a significant level of Internet censorship, but does not adopt the effect we describe in any significant way, and instead continues to have a notably high degree of foreign connectivity with $2,733$ first degree neighbours from $1,278$ ASes, albeit with a significant number of these existing under state-owned entities, such as \emph{Rostelecom} and its subsidiaries.

\begin{figure}[htp]
    \centering
    \includegraphics[width=\linewidth]{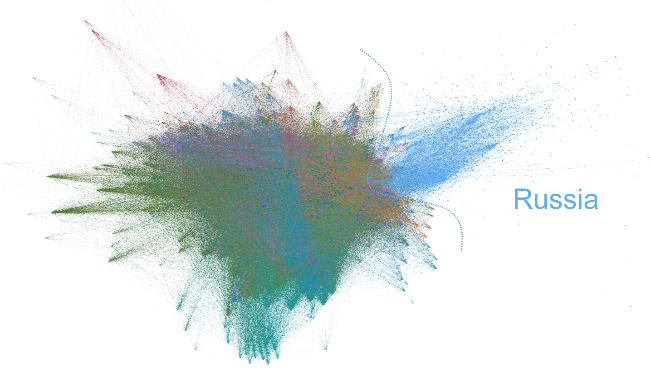}
    \caption{\textbf{International connectivity of Russia.} The figure shows Russia and its first degree of foreign neighbours. Russia is unusual relative to other authoritarian countries in having tighter interconnectivity, but press reports have suggested Internet domestic operators have capability to disconnect upon government instruction \cite{marrow_russia_2021}.}\label{country:russia}
\end{figure}

Despite this, the country does maintain a strong domestic network (with $5,078$ ASes) with a prominent tiering system, with many layers of downstream ASes from first degree foreign neighbouring nodes. This is demonstrated in Figure \ref{country:russia}.

Our work focuses on the architectural changes arising from geopolitical influences, however our work does not account for a small number of states adopting the approach taken by Russia, wherein state regulation supersedes architectural approaches to censorship, exemplified by domestic work on ``Runet'' \cite{marrow_russia_2021}.

\subsection{Geopolitical Fragmentation}

\begin{figure}[htp]
    \centering
    \begin{subfigure}[b]{0.48\linewidth}
        \centering
        \includegraphics[width=\textwidth]{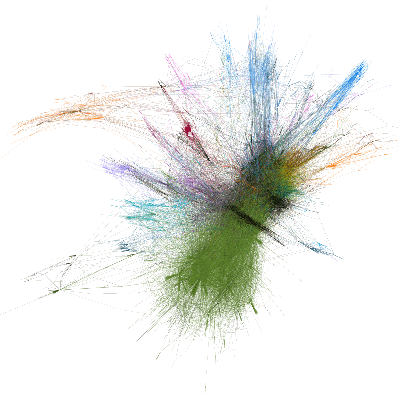}
        \caption{\small 1 June 2005}
        \label{fig:internet-development-2005}
    \end{subfigure}
    \hfill
    \begin{subfigure}[b]{0.48\linewidth}  
        \centering 
        \includegraphics[width=\textwidth]{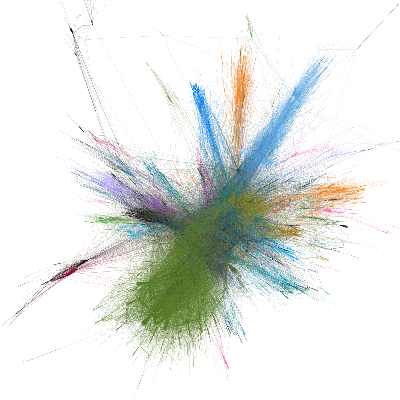}
        \caption{\small 1 June 2010}
        \label{fig:internet-development-2010}
    \end{subfigure}
    \vskip\baselineskip
    \begin{subfigure}[b]{0.48\linewidth}   
        \centering 
        \includegraphics[width=\linewidth]{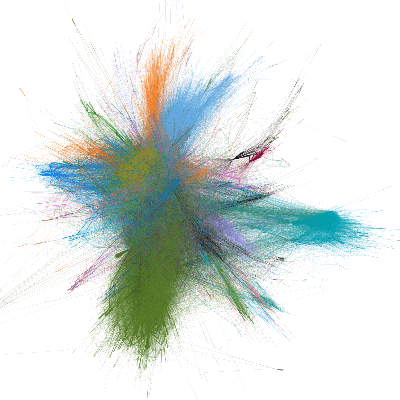}
        \caption{\small 1 June 2015}
        \label{fig:internet-development-2015}
    \end{subfigure}
    \hfill
    \begin{subfigure}[b]{0.48\linewidth}   
        \centering 
        \includegraphics[width=\textwidth]{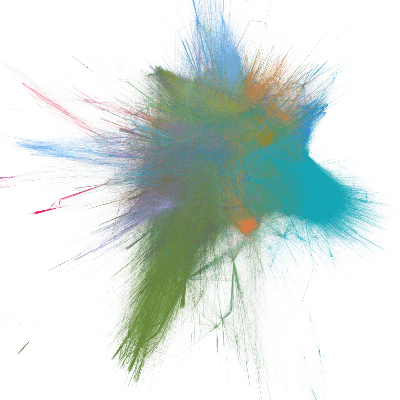}
        \caption{\small 1 June 2020}
        \label{fig:internet-development-2020}
    \end{subfigure}
    \caption{In this figure we observe the change in AS interconnection over 5-year intervals, starting in 2005. Although higher degrees of central interconnectivity emerge as the number of ASes and peerings increase, fragments are seen moving further away from the graph centre over time. Colours are consistent with those shown in earlier charts. The United States (in the bottom-left) is most prominent as a result of unusually high domestic connectivity and AS ownership. Countries exercising higher levels of Internet censorship are more greatly repelled, most clearly shown around the left and top edges of each graph.}
    \label{fig:internet-development}
\end{figure}

A final finding in our work is that of continued Internet fragmentation at a geopolitical level. In Figure \ref{fig:internet-development}, we show our force-directed topology over a 15-year period from June 2005 to June 2020, with an observable difference in topological shape, and a clear divergence of an increasing number of states from the centre point.

These findings are also evident statistically. Our data shows a $251\%$ increase in the number of observed ASes between 2005 and 2020 alongside an $879\%$ increase in the number of detected peerings, as well as a substantial increase in the mean degree of ASes (from $4.858$ in 2005 to $13.56$ in 2020). However, it is notable that this substantial increase in peerings has not been reflected in a statistically significant increase in the clustering coefficient -- meaning that, at best, the overall connectivity of the Internet's ASes has not improved, despite a substantial increase in the degree of the top $10\%$ of ASes (mostly tier-1/carrier providers) in the same period.

The average path length, despite the growing degree of carriers, also increases substantially from $3.7$ to $12.0$ over the 15-year period, further supporting our finding. Additionally, our metric considering the cumulative downstream burden demonstrates an increase in funnelling over time, with a mean of $0.491$ in 2005 compared with $0.359$ in 2020 (where increased funnelling results in a lower value).

The increase in funnelling, which we have shown to be correlated with an increase in Internet censorship effort, together with analysis of the characteristics of the Internet topology graph, enable us to demonstrate that the Internet is fragmenting over time.

\subsection{Limitations}

The most significant limitation of our work is its utilisation of BGP data from a set of vantage points that have little physical coverage of the continent of Africa. We do use a small number of route collectors in parts of south and western Africa, however coverage in some states is likely to be poor. This limitation, albeit notable, is shared by almost all other work analysing BGP data, and our effort to include additional data from an increased number of sources likely means our work has improved coverage when compared to others.

Secondly, the advertised \emph{peer\_ip} attributes in BGP data may not present complete coverage of the PoPs of an AS, and thus we may not be able to comprehensively determine all of the physical locations of an AS. We believe that we partially mitigate this with the scale of our data and by making our conclusions utilising a selection of approaches and metrics not dependent on this data alone.

Finally, our approach to handling potential outdated elements of the \emph{State-Owned ASes} dataset \cite{carisimo_identifying_2021} may result in a small number of false-positives or false-negatives. We only use this data to indicate key trends, and therefore its impact on our conclusions is likely to be negligible.
\section{Ethical Considerations}
\label{sec:ethics}

Despite making Internet measurements about censorship, which in some cases may raise significant ethical questions \cite{jones_ethical_2015}, our work does not involve human subjects, measurements taken on-location, or measurements directly related to censorship. Instead, we utilise publicly available data which is revealing of Internet topology, and therefore this work does not raise any ethical issues. In addition, we utilise data on Internet censorship from the V-Dem v13 dataset \cite{noauthor_v-dem_nodate, pemstein_v-dem_2022}, where we believe the authors have taken all reasonable steps to minimise the risk of potential harm.
\section{Conclusion}
\label{sec:conclusion}

Our work in this paper presents a novel approach to identifying geopolitical restrictions in Internet architecture. We first capture the Internet topology with completeness $7.1\%$ greater than existing state-of-the-art approaches \cite{wolfson_caidas_2021}, and use it to generate a graph-based topology model with which we supplement the routing data with relevant metadata related to the registration of Internet resources, physical location, and owner.

We then perform a latitudinal study of ASes at a nation-state level, presenting key trends analysed further through graph metrics. We then present our novel metric for measuring \emph{funnelling}, a phenomenon that we discover through the analysis of the Internet architecture in many countries with higher levels of Internet censorship.

Finally, through a limited longitudinal study, we show that the Internet is experiencing increased geopolitical fragmentation over time. We accompany this conclusion with a combination of graphical demonstrations based on our collected topology data\footnote{We intend to publish the tool for generating this topology data in a GitHub repository.}, as well as the use of our novel approach to detect funnelling.

\textbf{Future Work.} In the next iteration of our work, we hope to improve the visibility of interconnections within lesser observed regions, with a particular focus on Africa, as well as improving our inference-based approach to detect a greater number of private interconnections, which at present are largely found through traceroute data.

\end{document}